\RequirePackage{fix-cm}
\documentclass[smallcondensed]{svjour3}     

\smartqed  
\usepackage{graphicx}
\usepackage{etoolbox}

\usepackage{dcolumn}%
\institute{J. Raj \at
              Jagiellonian University \\
              Tel.: +48-57-6842258\\
              \email{juhi.raj@doctoral.uj.edu.pl}           
}

\begin{document}
\title{A feasibility study of the time reversal violation test based on polarization of annihilation photons from the decay of ortho-Positronium with the J-PET detector}
\author{J.~Raj$^1$ \and A.~Gajos$^1$ \and C.~Curceanu$^2$ \and E.~Czerwi\'nski$^1$ \and K.~Dulski$^1$ \and  M.~Gorgol$^3$ \and N.~Gupta-Sharma$^1$ \and B.C.~Hiesmayr$^4$ \and B.~Jasi\'nska$^3$ \and K.~Kacprzak$^1$ \and \L.~Kap\l on$^1$\and D.~Kisielewska$^1$ \and K.~Klimaszewski$^5$ \and G.~Korcyl$^1$ \and P.~Kowalski $^5$\and T.~Kozik$^1$ 
\and N.~Krawczyk$^1$\and W.~Krzemie\'n$^5$ \and E.~Kubicz$^1$ \and M.~Mohammed$^7$ \and  Sz.~Nied\'zwiecki$^1$ \and M.~Pa\l ka$^1$ \and M.~Pawlik-Nied\'zwiecka$^1$ \and L.~Raczy\'nski$^5$ \and K.~Rakoczy$^1$ \and Z.~Rudy$^1$ \and S.~Sharma$^1$ \and Shivani$^1$ \and R.Y.~Shopa$^5$ \and M.~Silarski$^1$ \and M.~Skurzok$^1$ \and W.~Wi\'slicki$^5$ \and B.~Zgardzi\'nska$^3$ \and  P.~Moskal$^1$} 
\maketitle
\textit{1. Faculty of Physics, Astronomy and Applied Computer Science, Jagiellonian University, 30-348 Cracow, Poland } \newline
\textit{2. National Institute of Nuclear Physics, Laboratori Nazionali di Frascati, 00044, Frascati, Italy} \newline
\textit{3. Department of Nuclear Methods, Institute of Physics, Maria Curie-Sk\l odowska University, 20-031, Lublin, Poland. }\newline
\textit{4. Faculty of Physics, University of Vienna, 1090, Vienna, Austria} \newline
\textit{5. \'Swierk Department of Complex Systems, National Centre for Nuclear Research, 05-400 Otwock-Świerk, Poland } \newline
\textit{6. Department of Physics, College of Education for Pure Sciences, University of Mosul, Mosul, Iraq} \newline

\begin{abstract}
The Jagiellonian Positron Emission Tomograph (J-PET) is a novel device being developed at Jagiellonian University in Krakow, Poland based on organic scintillators.
J-PET is an axially symmetric and high acceptance scanner that can be used as a multi-purpose detector system.
It is well suited to pursue tests of discrete symmetries in decays of positronium in addition to medical imaging. 
J-PET enables the measurement of both momenta and the polarization vectors of annihilation photons. The latter is a unique feature of the J-PET detector which allows the study of time reversal symmetry violation
operator which can be constructed solely from the annihilation photons momenta before and after the scattering in the detector.
\keywords{J-PET Detector \and Time Reversal Symmetry \and ortho-Positronium \and Discrete Symmetries \and Momenta \and Polarization \and Annihilation photons}
\end{abstract}

\section{Introduction}
\label{intro}
{  
Testing whether Nature is invariant under time reversal
is one of the most curious challenges in physics.
Although, one can reverse the direction of motion in space, one cannot reverse the direction of elapsing time. 
Therefore, in order to test the time reversal violation, we investigate the expectation values for the T-odd-symmetry-operators. 
To date,
time reversal symmetry violation has not been observed in purely leptonic systems. 
The best experimental upper limits for CP and CPT (C-Charge Conjugation, P-Parity and T-Time) symmetries violation in positronium decay are set to 0.3$\times$10\textsuperscript{-3}~\cite{Ref1,Ref2,Ref3}. 
According to the Standard Model predictions, photon-photon interaction or weak interaction can mimic the symmetry violation in the order of 10\textsuperscript{-9}(photon-photon interaction) and 10\textsuperscript{-13} (weak interactions), respectively~\cite{Ref4,Ref5,Ref6,Ref7}.
There is about 7 orders of magnitude difference between the present experimental upper limit and the standard model predictions.
So far, discrete symmetries were proposed to be tested with the ortho-positronium (o-Ps) system by determining the non-zero expectation values of the operators constructed from the spin (\vec{S}) of the o-Ps and the momentum vectors of the annihilation photons \vec{k_1}, \vec{k_2} and \vec{k_3} as mentioned in Table 1. 

\begin{table}[h!]
\centering
\caption{Discrete symmetry odd-operators using spin orientation of the o-Ps and momentum directions of the annihilation photons}
\begin{tabular}{|l|l|l|l|l|l|}
\hline
\textbf{Operator}          & \textbf{C} & \textbf{P} & \textbf{T} & \textbf{CP} & \textbf{CPT} \\ 
\hline
$\vec{S}\cdot\vec{k_1}$        & +  & $-$  & +  &  $-$   & $-$  \\
$\vec{S}\cdot(\vec{k_1}\times\vec{k_2})$ & +  & +  &  $-$  & +   & $-$ \\
$(\vec{S}\cdot\vec{k_1})\cdot(\vec{S}\cdot(\vec{k_1}\times\vec{k_2}))$ & +  & $-$  &  $-$  & $-$   & + \\
\hline
\end{tabular}
\end{table}
In this article we explore feasibility of applying a new method of studying the fundamental discrete symmetry violation in a purely leptonic system proposed in Ref.~\cite{Ref8}. The novelty of the method is constituted by using of the discrete symmetry odd-operators constructed with the polarization (\vec{\epsilon_i}) of the annihilation photons originating from the decay of o-Ps atoms, as listed in Table 2~\cite{Ref8}.  
The observation of non-zero expectation values of these operators would imply non-invariance of these symmetries for which the given operator is odd
(marked $"-"$ in Table 1 and Table 2).

\begin{table}[h!]
\centering
\caption{Discrete symmetry odd-operators using spin orientation of the o-Ps as well as polarization and momentum directions of the annihilation photons}
\begin{tabular}{|l|l|l|l|l|l|}
\hline
\textbf{Operator}          & \textbf{C} & \textbf{P} & \textbf{T} & \textbf{CP} & \textbf{CPT} \\ 
\hline
$\vec{\epsilon_1}\cdot\vec{k_2}$        & +  & $-$  & $-$  & $-$ & +    \\
$\vec{S}\cdot\vec{\epsilon_1}$         & +  & +  & $-$  &  +  & $-$    \\
$\vec{S}\cdot(\vec{k_2}\times\vec{\epsilon_2})$ & +  & $-$  & +  & $-$   & $-$  \\
\hline
\end{tabular}
\end{table}

\section{Jagiellonian - Positron Emission Tomograph}
\label{jpet}
Positron emission tomography (PET) is a non-invasive technique used in the diagnosis of various types of tumors at the cellular level. All commercially available PET-scanners utilize relatively expensive crystal detectors for the detection of annihilation photons~\cite{Ref16,Ref17}.  

The Jagiellonian - Positron Emission Tomograph (J-PET) is the first PET-scanner constructed using plastic scintillator strips to make the tomograph cost effective and portable~\cite{Ref9,Ref10,Ref13}. One of the unique features of the J-PET detector is its ability to measure  polarization of the annihilation photons.
The J-PET detector consists of 192 plastic scintillator strips (EJ-230) of dimensions $500 \times 19 \times 7~mm^3$ each, forming three concentric layers (48 modules on radius 425~mm, 48 modules on radius 467.5~mm and 96 modules on radius 575~mm) (Fig.~\ref{fig:1})\cite{Ref13}. 
\vspace{-0.5cm}
\begin{figure*}[h!]
\includegraphics[width=0.5\textwidth]{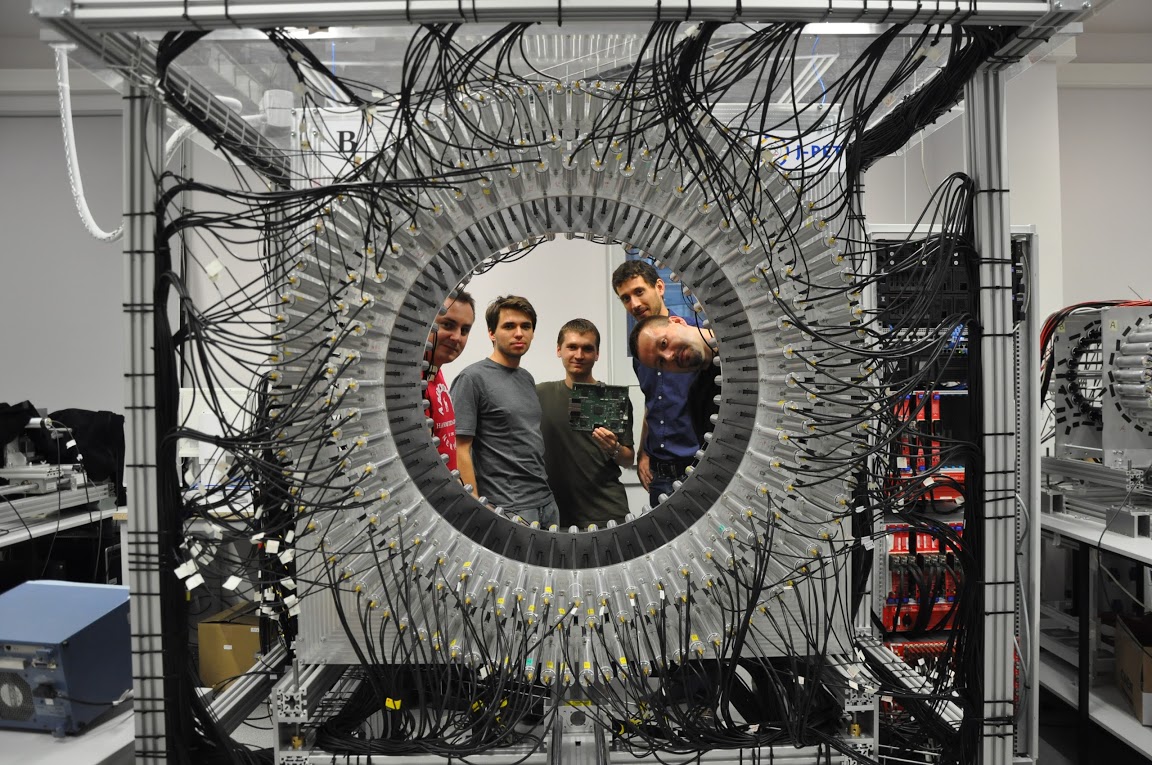}
\includegraphics[width=0.5\textwidth]{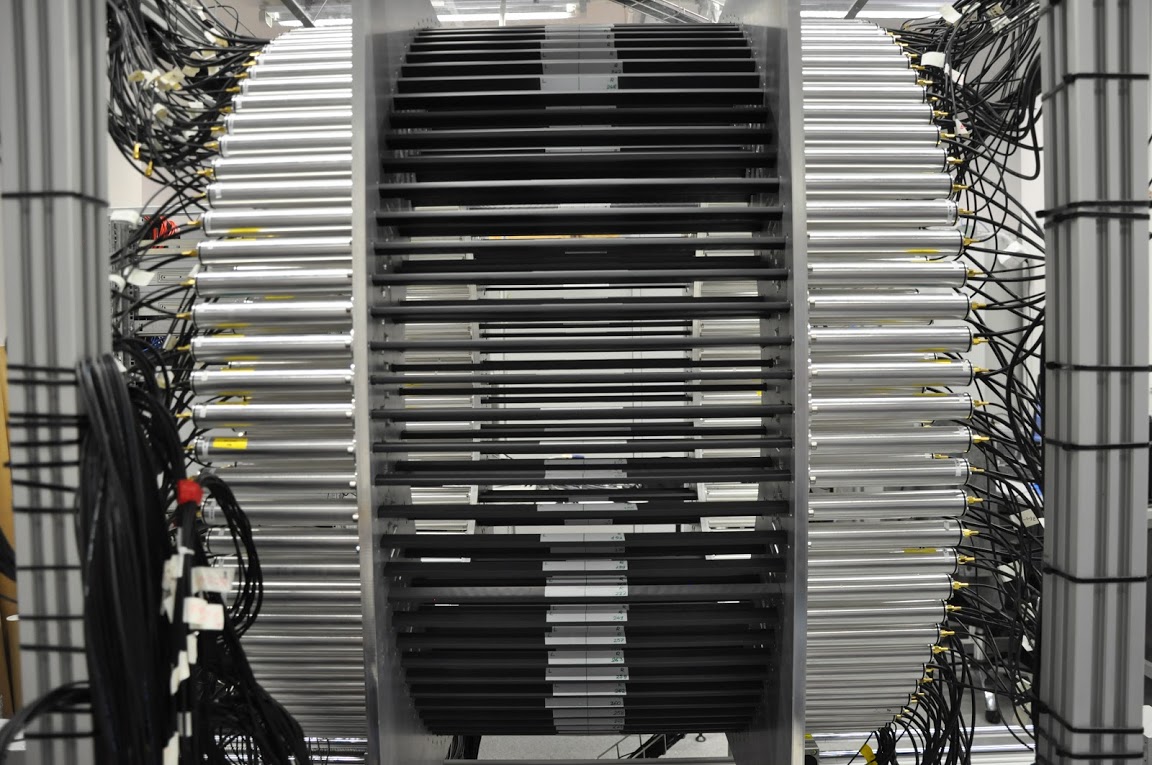}
\caption{(Left) Front view of the J-PET detector with a few members of the J-PET group. (Right) Side view of the J-PET detector, where, the black strips are the scintillators wrapped in light protective foil. The silver tubes are housings of  photomultipliers which are connected at either ends of the scintillator.}
\label{fig:1}
\vspace{-0.5cm}
\end{figure*}
Each scintillator in the J-PET scanner is optically connected at two ends to Hamamatsu R9800 vacuum tube photomultipliers~\cite{Ref10}.
This results in 384 analog channels that are processed by a fully equipped trigger-less Data Acquisition System (DAQ) together with the readout mechanism~\cite{Ref14,Ref18}.
The readout chain consists of Front-End Electronics (FEE) such as Time-to-Digital or Analog-to-Digital Converters (TDCs or ADCs), data collectors and storage~\cite{Ref14,Ref18}. 

The J-PET collaboration has developed a dedicated J-PET analysis framework which is a highly flexible, ROOT-based software package which aids the reconstruction and calibration procedures for the tomograph. 
The J-PET analysis framework stands as a backbone to analyze the collected data to produce the results presented in this article \cite{Ref15}. 

The J-PET detector, together with the trigger-less DAQ system constitutes an efficient photon detector with high timing properties \cite{Ref13}.
This allows us to investigate the fundamental discrete symmetries in the purely leptonic sector \cite{Ref8}.

\section{Measurement Technique}
\subsection{Production of positronium}
In order to produce positronium atoms, a point-like $^{22}$Na source is placed in the center of the detector and is surrounded with XAD-4 porous polymer~\cite{Ref19}. The porous polymer enhances the production of positronium atoms.
\begin{figure*}[h!]
\includegraphics[width=0.51\textwidth]{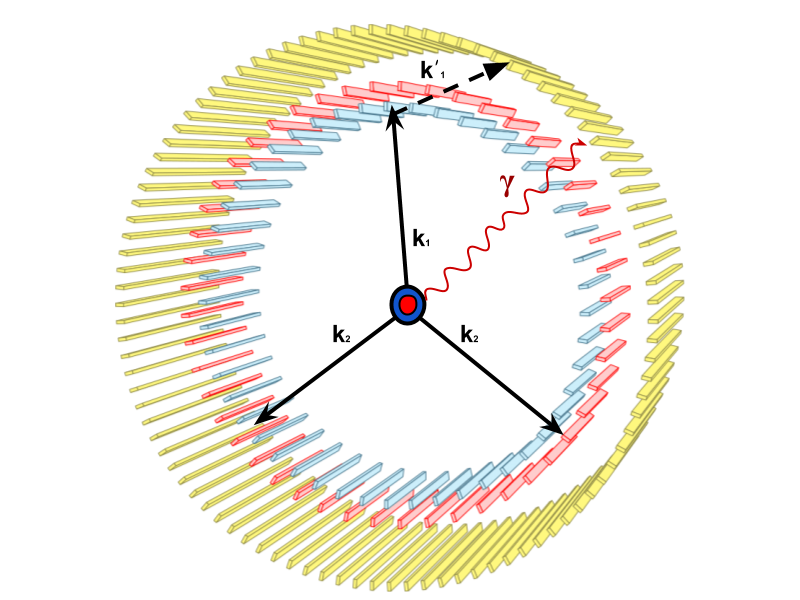}
\centering
\caption{Schematic of the J-PET detector with three distinguishable layers of scintillators at blue, red and yellow concentric cylinders. A positron source (red) placed in the center, covered in XAD-4 porous polymer (blue). The superimposed arrows indicate gamma photon originating from the de-excitation of $^{22}Ne^*$ ($\gamma$), annihilation photons from ortho-positronium decay ($k_1$, $k_2$ and $k_3$), and scattered photon ($k'_1$).   
\label{fig:2} 
}
\end{figure*} \newline
The positrons emitted from the source 
($\textsuperscript{22}Na \rightarrow \textsuperscript{22}Ne^{*} + e^{+} + \textit{v}_e$),
interact with the electrons in the XAD-4 porous polymer producing a meta-stable triplet state, ortho-Positronium (o-Ps) which predominantly decays into three photons due to charge conjugation symmetry conservation \cite{Ref7}:
$e^{+} + e^{-} \rightarrow \textit{o-Ps} \rightarrow 3\gamma$.  
Meanwhile, the excited $^{22}Ne^{*}$ de-excites emitting gamma quantum with an energy of 1274~keV  
 ($\textsuperscript{22}Ne^{*} \rightarrow \textsuperscript{22}Ne + \gamma_{(1274\ keV)}$). 
\subsection{Selection of o-Ps decay events}
\vspace{-1cm}
\begin{figure*}[h!]
\includegraphics[width=0.53\textwidth]{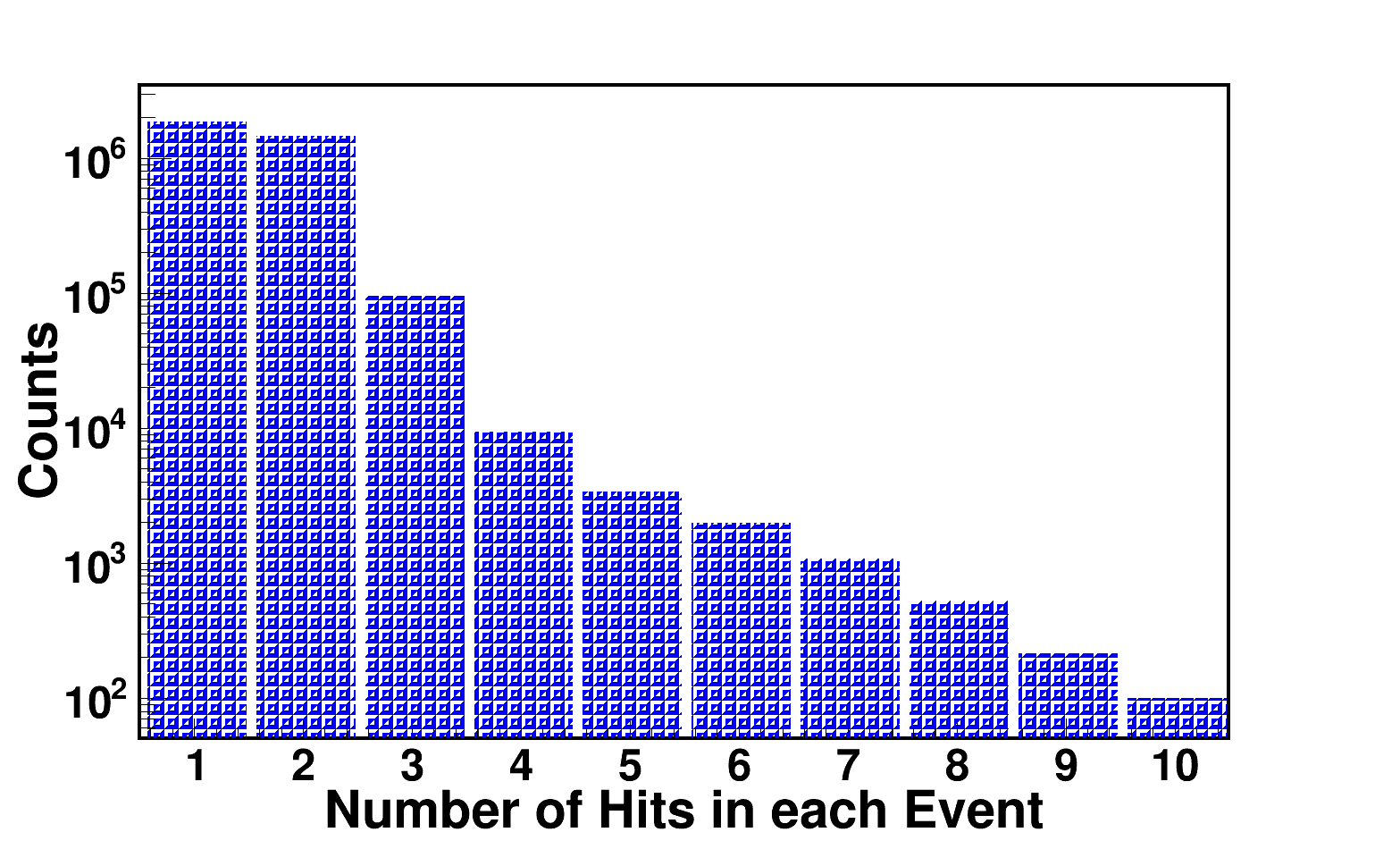}
\includegraphics[width=0.53\textwidth]{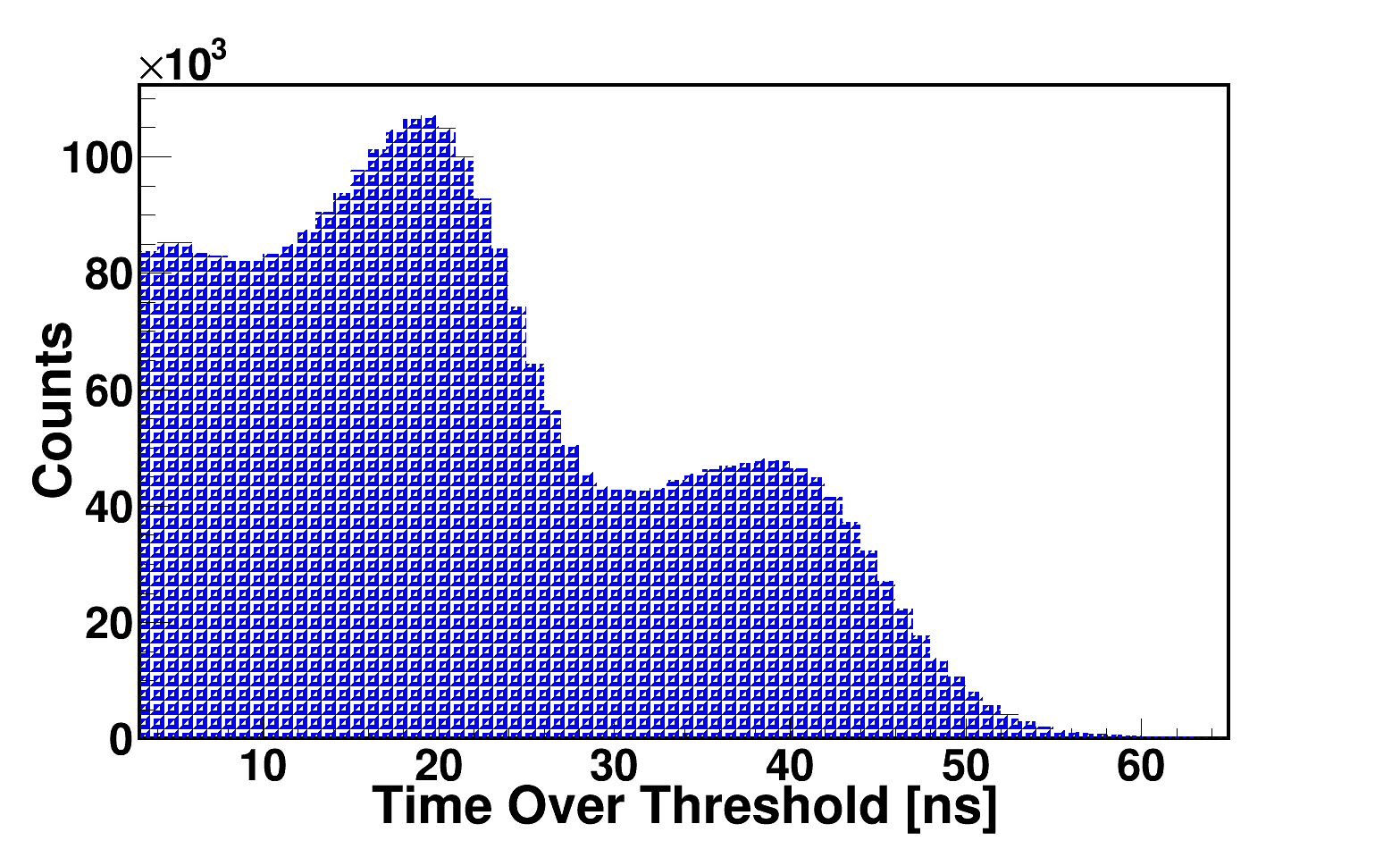}
\caption{(Left) The distribution of number of gamma interactions recorded in each event. (Right) Time over threshold of photomultiplier signals, used as a measure of the energy deposited in the scintillator strips. 
\label{fig:3}
}
\end{figure*}
\vspace{-0.4cm}
For the study of time reversal symmetry in o-Ps we are interested in events with five interactions which are due to: 
one de-excitation photon,
three primary annihilation photons, 
and one secondary scattered photon 
($k'_1$), as indicated in 
Fig.~\ref{fig:2}.
The signal (hit) multiplicity distribution presented in the left panel of 
Fig.~\ref{fig:3} shows that the requirement of five hits in one event reduces the measured data sample by a factor of about
$10^3$.
The de-excitation photon is identified using the time-over-threshold (TOT) measurement
~\cite{Ref20} 
which is related to the energy 
deposited in the scintillator. Right panel of 
Fig.~\ref{fig:3} presents TOT distribution where one can clearly recognize Compton spectra from 511~keV and 1274~keV gamma photons. The de-excitation photon (1274 keV) may be selected with the efficiency of about 66\% when requiring TOT larger than 30~ns. 

After selection of the de-excitation photon, we need to identify events in which annihilation photons originate from the decay of ortho-positronium, and relate the primary annihilation photon to its corresponding secondary scattered photon.

From kinematics we infer that the sum of the two smallest relative azimuthal angles between the registered annihilation photons for \textit{o-Ps}$\rightarrow 3\gamma$ must be greater than 180$^{\circ}$~\cite{Ref11,Ref12}.
Therefore, the events corresponding to the decay of \textit{o-Ps}$\rightarrow 3\gamma$ lie at the right side of the band at 180$^{\circ}$ (Fig.~\ref{fig:4}(Left)) \cite{Ref11,Ref12}.
\vspace{-0.5cm}
\begin{figure*}[h!]
\includegraphics*[width=0.51\textwidth]{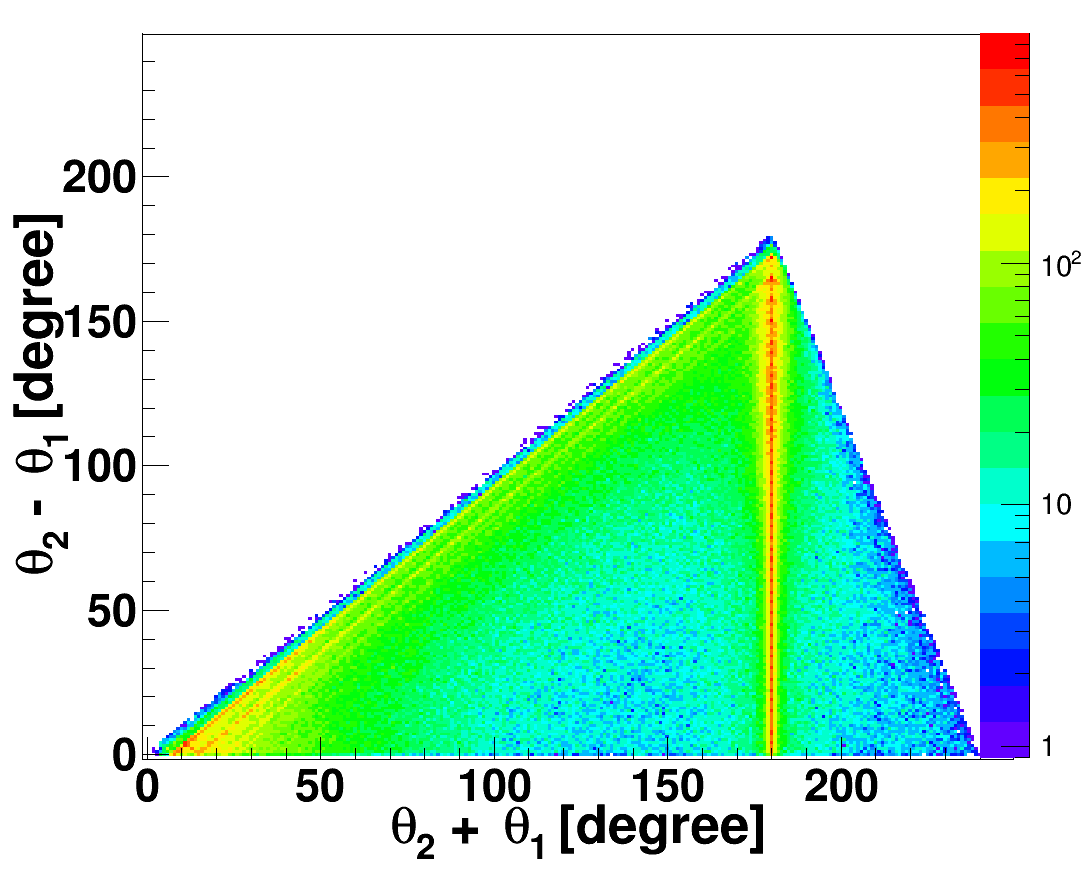}
\includegraphics*[width=0.51\textwidth]{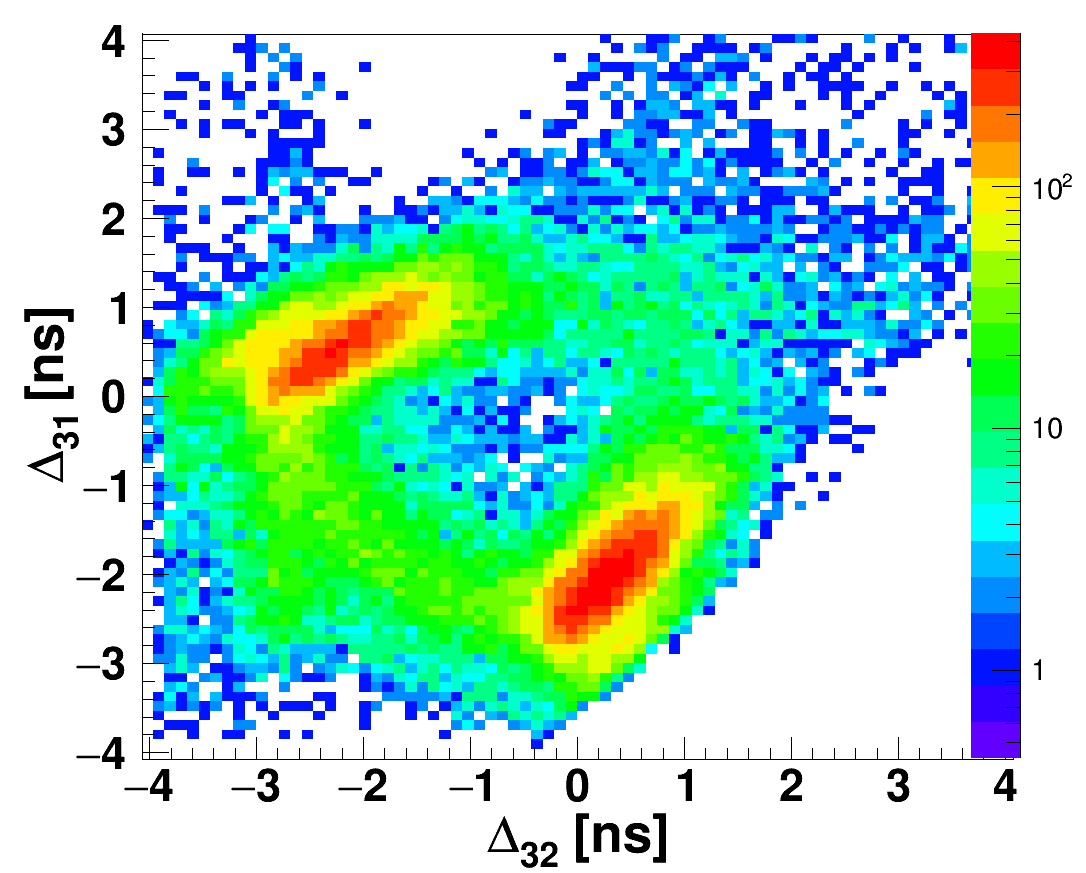}
\caption{(Left) Relation between the sum and difference of two smallest relative azimuthal angles ($\theta_1$ and $\theta_2$ (Figure. 2-Left)) of the three interacting annihilation photons. (Right) Distribution of $\Delta_{31}$ vs $\Delta_{32}$ as explained in the text.
\label{fig:4}
}
\end{figure*}
\vspace{-0.5cm}
The signals corresponding to the annihilation photons and secondary scattered photon may be identified after ordering the hits in time.
The signal from the scattered photon should appear as the last one. 
As a last step we need to assign the scattered photon to its primary photon. For this purpose we introduce a parameter $\Delta_{ij}~=~(\delta_M~-~\delta_C)$, where, $\delta_M$ and $\delta_C$ are the measured time difference and calculated time of flight between the $i^{th}$ and $j^{th}$ interaction, respectively. 
Therefore, $\Delta_{ij}$ should be equal to zero in case if the $j^{th}$ signal is due to the $i^{th}$ scattered photon. 
Right panel of Figure~\ref{fig:4} shows an exemplary spectrum allowing to relate the primary annihilation photon to its corresponding secondary photon.

Finally, the direction of polarization ($\vec{\epsilon_1}$) of the 1$^{st}$ (most energetic) annihilation photon is estimated as the cross-product of the momentum of the primary and secondary photons (\vec{k_1}x\vec{k'_1})~\cite{Ref8},
and is used in the determination  of the expectation value of the time symmetry odd-operator $\vec{\epsilon_1} \cdot \vec{k_2}$.

\section{Conclusion}
Positronium is an excellent laboratory enabling studies of many interesting phenomena~\cite{Ref21} such as e.g. gravitation of antimatter~\cite{Ref21}, search for mirror photons~\cite{Ref22,Ref23
}, quantum entanglement~\cite{Ref24,Nowakowski} and tests of discrete symmetries in the leptonic sector~\cite{Ref8}. 
So far, T-violation effects were not discovered
in a purely leptonic system e.g in the positronium system. 
The Jagiellonian-PET scanner offers high acceptance in the measurement of the direction of polarization ($\vec{\epsilon}$) of photons simultaneously with their momentum direction ($\vec{k}$).
In reference~\cite{Ref11} it was shown that the J-PET detector is capable of registration of $oPs\to 3\gamma$ decays with the rate ranging between 25 to 18300 events per second, depending on the configuration and amount of the used scintillator strips. In 2018 J-PET will be updated with 320 new scintillator strips leading to the registration rate of about 450 per second for the $oPs\to 3\gamma$ events.
A need of registration of the additional scattered photon for the studies of $\vec{\epsilon_1} \cdot \vec{k_2}$ operator will reduce this rate to about 30 events per second.  Thus during one year of the data collection (achievable in practice in about 3 years) it is expected to collect about $10^9$ events and hence it is expected to increase the sensitivity by about order of magnitude with respect to previous analyses based on about $10^7$ events~\cite{Ref2,Ref3}. 


\section{Acknowledgement}
This work was supported by The Polish National Center for Research and Development through grant INNOTECH-K1/IN1/64/, 159174/NCBR/12, the Foundation for Polish Science through the MPD and TEAM/2017-4/39 programmes, the National Science Centre of Poland through grants no.\ 2016/21/B/ST2/01222, 2017/25/N/NZ1/00861, the Ministry for Science and Higher Education through grants no. 6673/IA/SP/2016,
7150/E-338/SPUB/2017/1 and 7150/E-338/M/2017, and the Austrian Science Fund FWF-P26783.

\end{document}